\documentclass[aip,JCK]{revtex4-1}
\usepackage[dvips]{epsfig}
\usepackage{amsmath}
\usepackage{varwidth,xcolor}
\usepackage{graphicx}
\usepackage{subfig}
\usepackage{float}
\DeclareGraphicsExtensions{.pdf,.jpeg,.png,.eps,.jpg}
\begin{document}

\title{Entropic rectification and current inversion in a pulsating channel}

\author{M. Florencia Carusela}
\email{flor@ungs.edu.ar}
\affiliation{Instituto de Ciencias, Universidad Nacional de General Sarmiento, J.M.Gutierrez 1150, CP 1163, Los Polvorines, Buenos Aires, Argentina; \\ CONICET, Argentina}
\author{J. Miguel Rub\'i}
\affiliation{Departament de F\'isica Fonamental, Universitat de Barcelona, C/Mart\'i Franqu\`es 1, 08028, Barcelona, Spain}

\maketitle

\begin{quotation}

We show the existence of a resonant behavior of the current of Brownian particles confined in a pulsating  channel. The interplay between the periodic oscillations of the shape of the channel and a force applied along its axis leads to an increase of the particle current as a function of the noise level. A regime of current inversion is also observed for particular values of the oscillation frequency and the applied force. The model proposed to obtain these new behaviors of the current is based on the Fick-Jacobs equation in which the entropic barrier and the effective diffusion coefficient depend on time. The phenomenon observed could be used to optimize transport in microfluidic devices or biological channels.

\end{quotation}

\date{}

\section{Introduction}

Transport of particles through corrugated narrow channels is frequently found in physico-chemical and biological systems in which particles move along a main transport direction \cite{sugi,brun,kreu,siwy}.

Confinement changes significantly the transport properties of the particles and the energy conversion mechanisms \cite{rubi2,rubi4,rubi4b}.
The presence of boundaries yield a nontrivial contribution to the dynamics that manifests in the appearance of entropic forces in the Fick-Jacobs equation. It provides a very accurate description of entropic transport in 2D and 3D channels of varying cross-section. This equation is equivalent to a Smoluchowski equation in 1D dimension \cite{zwanzig,rubi1,rubi3,dad,dad2,dad3}.

In many of the studies carried out up to now, channels are considered as rigid structures as for kinetics of molecular motors and diffusion in zeolites \cite{transl1,transl2,transl3,zeol}, cell membrane channels or ion translocation through protein channel \cite{ion,cell1,cell2} . However in some cases the flexible nature of the channel plays a very important role in the transport properties, as in the case of vascular \cite{vascul1,vascul2} or peristaltic channels \cite{perist} and flexible DNA nanochannels \cite{dna}.

Our purpose in this article is to propose an entropic transport model to analyze the transport of particles under the influence of a driving force, through a pulsating channel whose walls undergo an oscillatory motion.  The variation in time of the shape of the channel makes that the entropic barrier depend on time.  
A resonant transport of particles in a channel with periodically oscillating walls was found when the force is perpendicular to the direction of the flow but diffusion is constant\cite{pavol,entropic,entropicw}. We propose to include the effect of the entropic barrier on the diffusion constant. Specifically we consider an external constant force acting in the direction of the flow to study the  rectification and current reversal (CR) phenomena. The last is not found when the force is perpendicular and the diffusion coefficient is assumed not affected by the entropic barrier \cite{entropic,entropicw}.
The dynamics of the particles can be controlled by the noise level, the oscillation frequency and the strength of the force. For some values of these quantities, the current exhibits a resonant behavior and an inversion of its direction. 

The article is organized as follows. In Sec.II, we introduce the entropic transport model and the effective quantities used to characterized the transport. The results obtained are presented in Sec.III. In Sec.IV, we present our main conclusions.

\section{The model}

\begin{figure}[H]
\centering
\includegraphics[width=8cm,height=6cm]{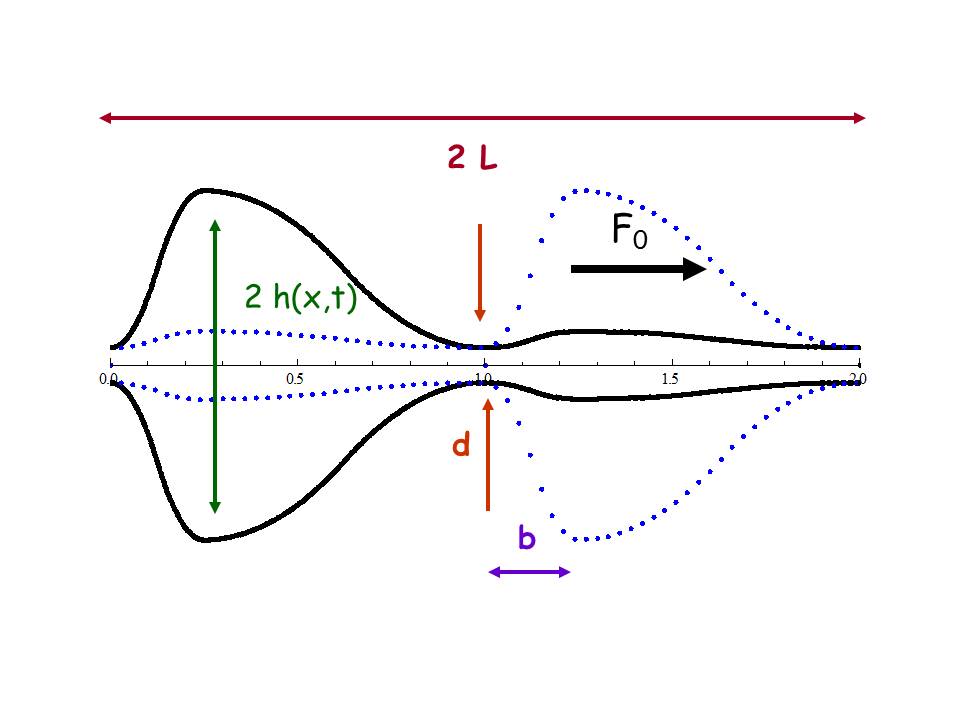}
\caption{Snapshot of a unit of the channel oscillating out of phase with period $T$, for two different times. The solid (black) line corresponds to $t=0$ and the dotted (blue) line corresponds to $t=T/2$.}
\label{fig:Figura1}
\end{figure}

We study the confined motion of $N$ non interacting Brownian particles through a two dimensional channel
which consists of units of length $2 L$, formed by two subunits of length $L$, as is shown in Fig.\ref{fig:Figura1}

The shape of the boundaries of the channel is periodically modulated in time. The height is given by:

\begin{equation}
h(x,t)= \left\{ \begin{array}{lcc}
             a_1(t) x^2 + \frac{d}{2} &   ;  &  0 \leq x \leq \frac{b}{2} \\
             \\ -a_1(t) (x-b)^2 + s(t)  &   ;  &  \frac{b}{2}< x  \leq  b\\
             \\ -a_2(t) (x-b)^2 + s(t)  &   ;  &  \frac{b}{2}< x  \leq \frac{L+b}{2}\\
						 \\ a_2(t) (x-L)^2 +   \frac{d}{2} &   ;  &  \frac{L+b}{2} < x \leq  L  
             \end{array}
   \right.
\end{equation}
where $b$ indicates the location of the point of maximun width and $d$ is the width of the bottleneck.   

The time dependent coefficients are $a_1(t)=\frac{2[s(t)-d]}{b ^2}$,  $a_2(t)=\frac{2[s(t)-d]}{(L-b)^2}$ and $s(t)=s_0+s_1 sin(\omega t+\Phi) $. The parameters are kept fixed with values that guarantee the asymmetry of the unit cell.  The phase difference between adjacent subunits of each unit is given by:

\begin{equation}
\Phi= \left\{ \begin{array}{lcc}
             0 &   ;  &   x \in [0,L]\\
						 \\ (0,\pi]   &   ;  &  x \in  (L ,2 L]
             \end{array}
   \right.
\end{equation}

Consecutive subunits shrink and enlarge alternatively, corresponding $\Phi=\pi$ in order to preserve a constant volume.

\begin{figure}[H]
\centering
\includegraphics[width=7cm,height=5cm]{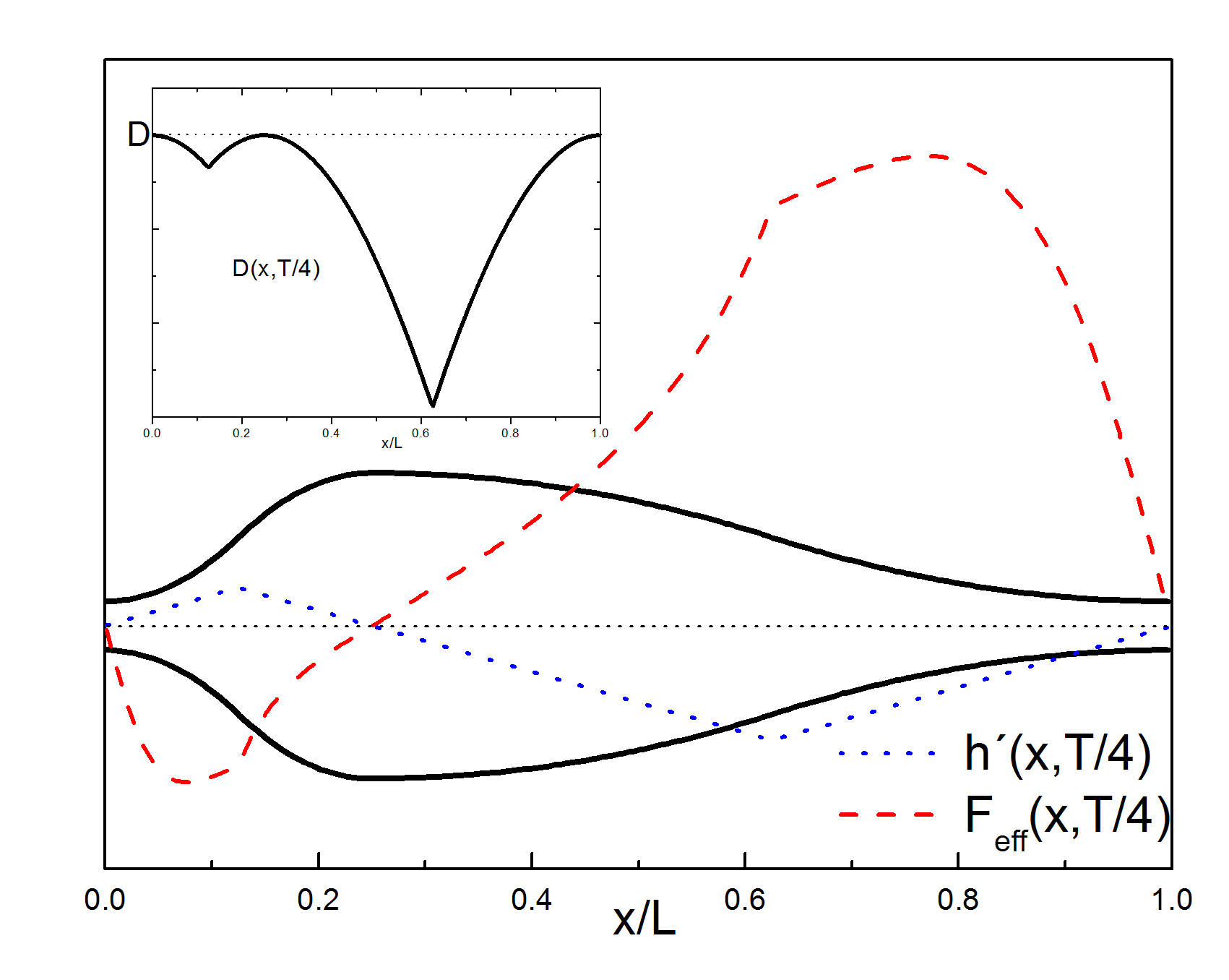}
\caption{Schematic illustration of $F_{eff}(x,t)$ (red dashed line), $h'(x,t)$ (blue dotted line) and $D(x,t)$ (inset) in a subunit cell (black full line). Time is chosen generically as $t=T/4$. }
\label{esquemas}
\end{figure}

We will analyze the transport properties by means of the Fick-Jacobs equation, that governs the dynamics of the probability distribution of the ensemble of non-interacting Brownian particles.

\begin{equation}
\frac{\partial P(x,t)}{\partial t}= \frac{\partial }{\partial t} \left[ D(x,t) \frac{\partial P(x,t)}{\partial x} +\frac{D(x,t)}{k_B T} V'_{eff}(x,t) P(x,t) \right]
\label{FJ}
\end{equation}
Here $D(x,t)$ is an effective diffusion coefficient, that in our case is given by \cite{rubi3}

\begin{equation}
D(x,t)= \frac{D_0}{ (1+h'(x,t)^2)^{\left(1/3\right)} }
\label{dif}
\end{equation}
where $D_0=k_B T/\eta$ is the diffusion coefficient for an unbounded medium and 

\begin{equation}
F_{eff}(x,t)= -\frac{\partial A(x,t)}{\partial x}= F_0 + k_B T \frac{h'(x,t)}{h(x,t)}
\label{feff}
\end{equation}
is the effective force acting in the x-direction. The free energy due to energy and entropic barriers is $A(x,t) \doteq E - T S = -F_0 x -k_B T \ln h(x,t)$ (see Fig.\ref{esquemas}).

From Eq.\ref{FJ}, we indentify the instantaneous particle current
\begin{equation}
J(x,t)= - \left[ D(x,t) \frac{\partial P(x,t)}{\partial x)} -\frac{D(x,t)}{k_B T} F_{eff}(x,t) P(x,t) \right]
\label{J}
\end{equation}

The Fick-Jacobs equation assumes that the probability density reaches equilibrium in the transverse direction much faster than in the longitudinal one. This requirement is fulfilled if $\left| h'(x,t) \right| << 1$ for all times and positions.

For the sake of simplicity, we will use dimensionless quantities.
We scale the lengths with the unit length $2L$, times with the diffusion time $\tau_{dif}=L^2 \eta/ (k_B T)$ ($\eta$ is the friction coefficient), energies with $k_B T$, forces with units of $k_B T /L$ and currents with of $L/\tau_{dif}$ . 

A typical diffusion constant in colloids in aqueous solution is $D_0 \approx 10^{-12}m^2/s$; therefore one typically has Brownian time scales of $1 - 100$s, for particles of sizes from $10^{-3}$ to $1 \mu$m and velocities in the range $10^{-2} - 10^{-1} (\mu m)/s$ \cite{Dalle,Lopez}.

In our case the validity of the Fick-Jacobs approach requires that the dimensionless frequency $\omega$ has to be smaller than one, this implies modulations smaller than $10$ Hz approximately. This range is of the order of the ones used in recent experiments on transport of molecules in confined media subject to entropic barriers and to a driving force \cite{Lairez}.

\section{Numerical Results}

From the Fick-Jacobs equation, we obtain numerically the probability density $P(x,t)$ with periodic boundary conditions at $x=0$ and $x=1$.

We introduce the mean particle current as

\begin{equation}
j(t) = \int^{1}_{0} J(x,t)  dx
\label{Jt}
\end{equation}
with the probability current $J(x,t)$ given in Eq.(\ref{J}).  It can be written as \cite{Reimann}

\begin{eqnarray}
j(t) &=& \frac{d}{dt}\int^{z_0+1}_{z_0} P(x,t) x dx + J(z_0,t) \\
&=& \frac{d}{dt} <x(t)> + J(1,t)
\label{Jt2}
\end{eqnarray}
where $z_0$ is an arbitrary reference position. Here we choose $z_0=1$, that is the upper boundary of the unit where the initial ensemble of particles is located.

Eq. \ref{Jt2} indicates that the particle current is composed by the motion of the 'center of mass' (CM)  of the ensemble and the probability current evaluated at the reference point. Then we define the position

\begin{equation}
X(t) \equiv \int^{t_0}_{t} j(t') dt' =<x(t)>_{CM} +  \int^{t_0}_{t} J(z_0,t) dt'
\label{xt}
\end{equation}

where we have used that $<x(t_0)>=<x(t_0)>_{CM}$.

We now average the current in a period $T=2\pi / \omega$ as

\begin{equation}
J = \frac{1}{T}\int^{T+t_0}_{t_0} j(t) dx
\label{Jt3}
\end{equation}

and the time average of the mean position position

\begin{equation}
{\overline X} =\frac{1}{T}\int^{T+t_0}_{t_0} X(t) dt
\label{xm}
\end{equation}

\begin{figure}[H]
\centering
\includegraphics[width=8cm,height=6cm]{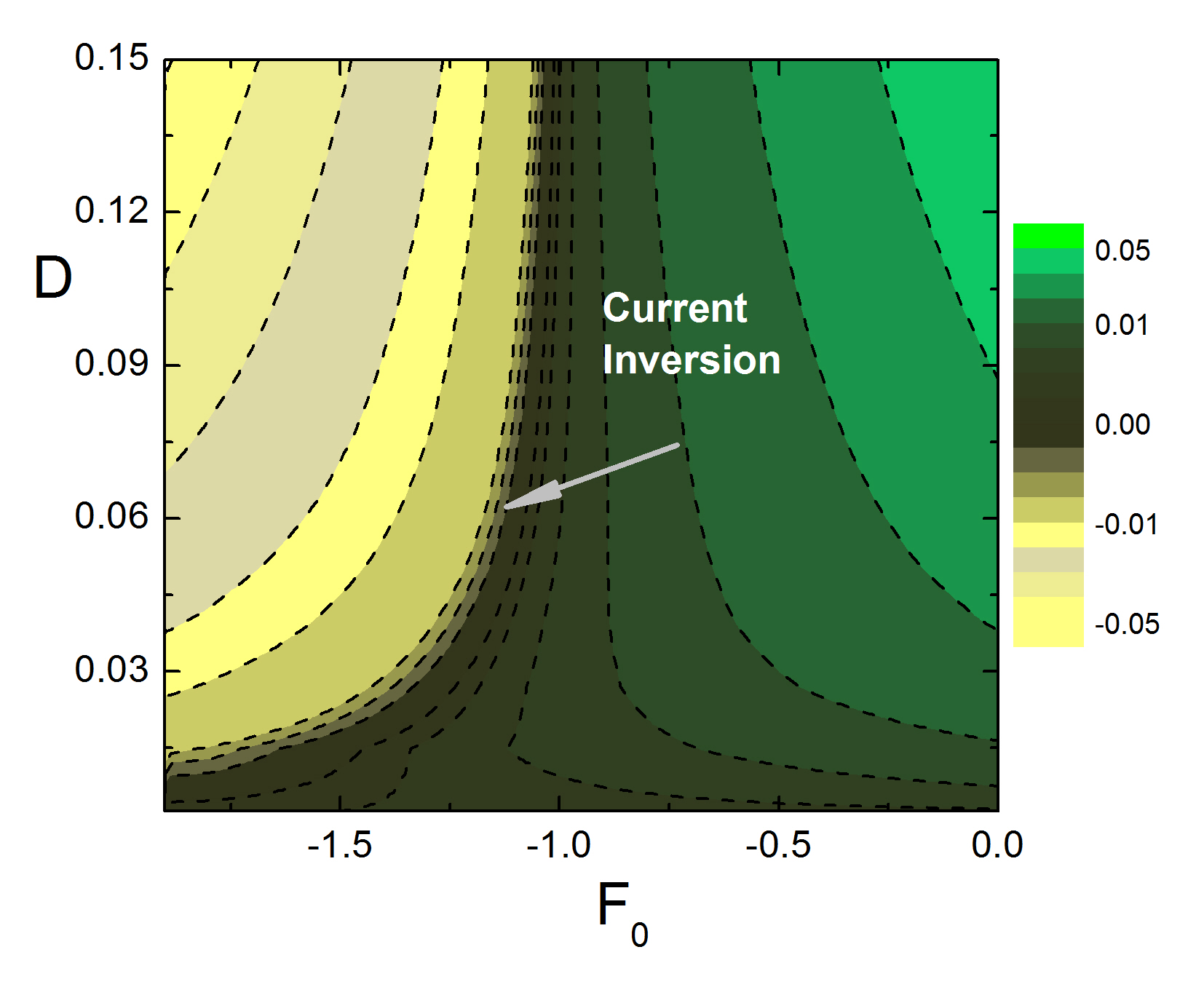}
\caption{Contour plot for J versus $D$ and $F_0$ for $\omega=0.5$.  The values of the parameters are: $s_0 = 0.45$, $s_1 = 0.35$, $b = 0.25$, $d = 0.15$, $L=1$.  Currents are given in units of $\mu m/s$}
\label{fig:conturs}
\end{figure}

\begin{figure}[h]
\begin{minipage}{0.5\textwidth}
\subfloat[]{\includegraphics[width=7cm,height=5cm]{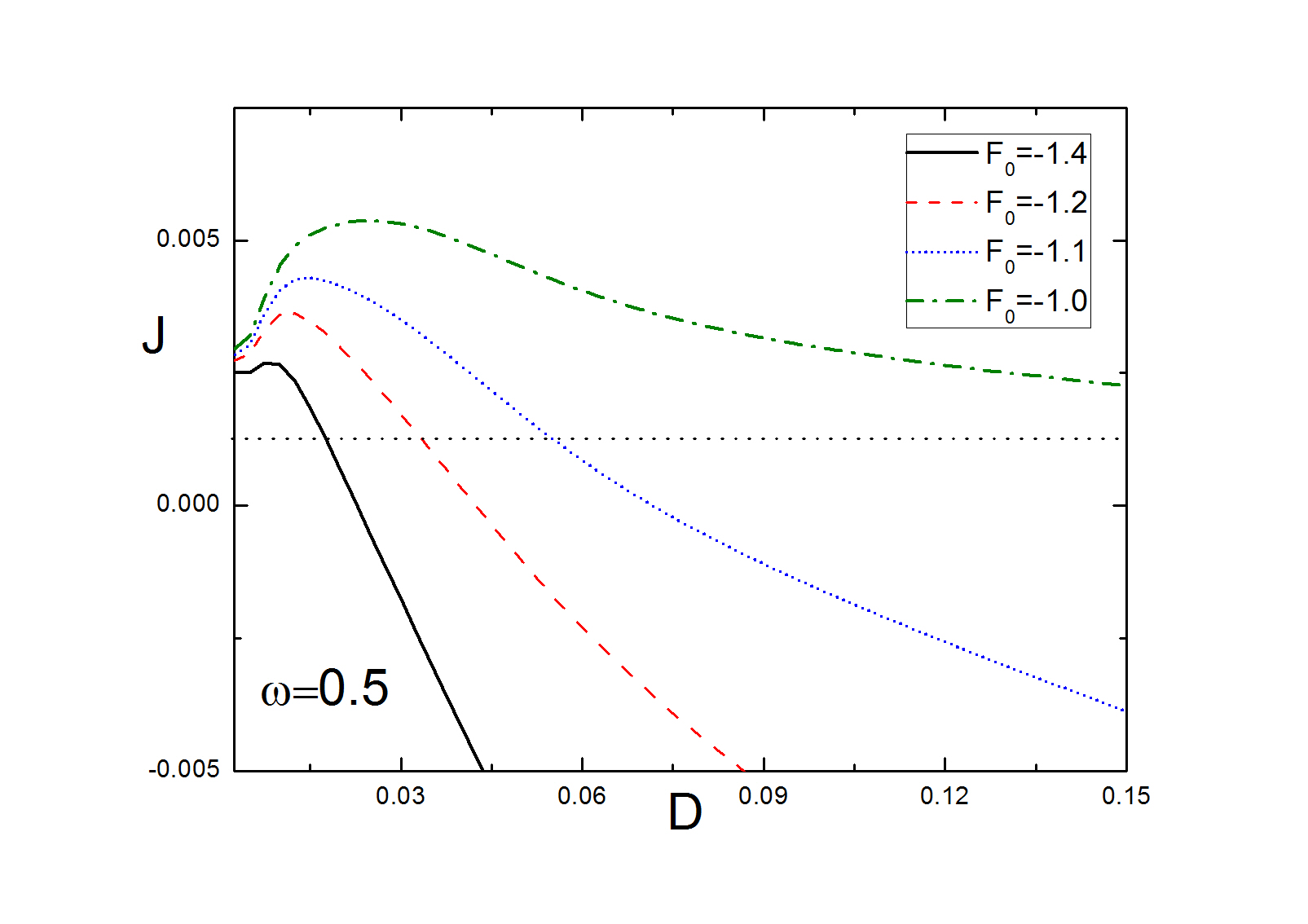}}
\end{minipage}%
\hfill
\begin{minipage}{0.5\textwidth}
\subfloat[]{\includegraphics[width=7cm,height=5cm]{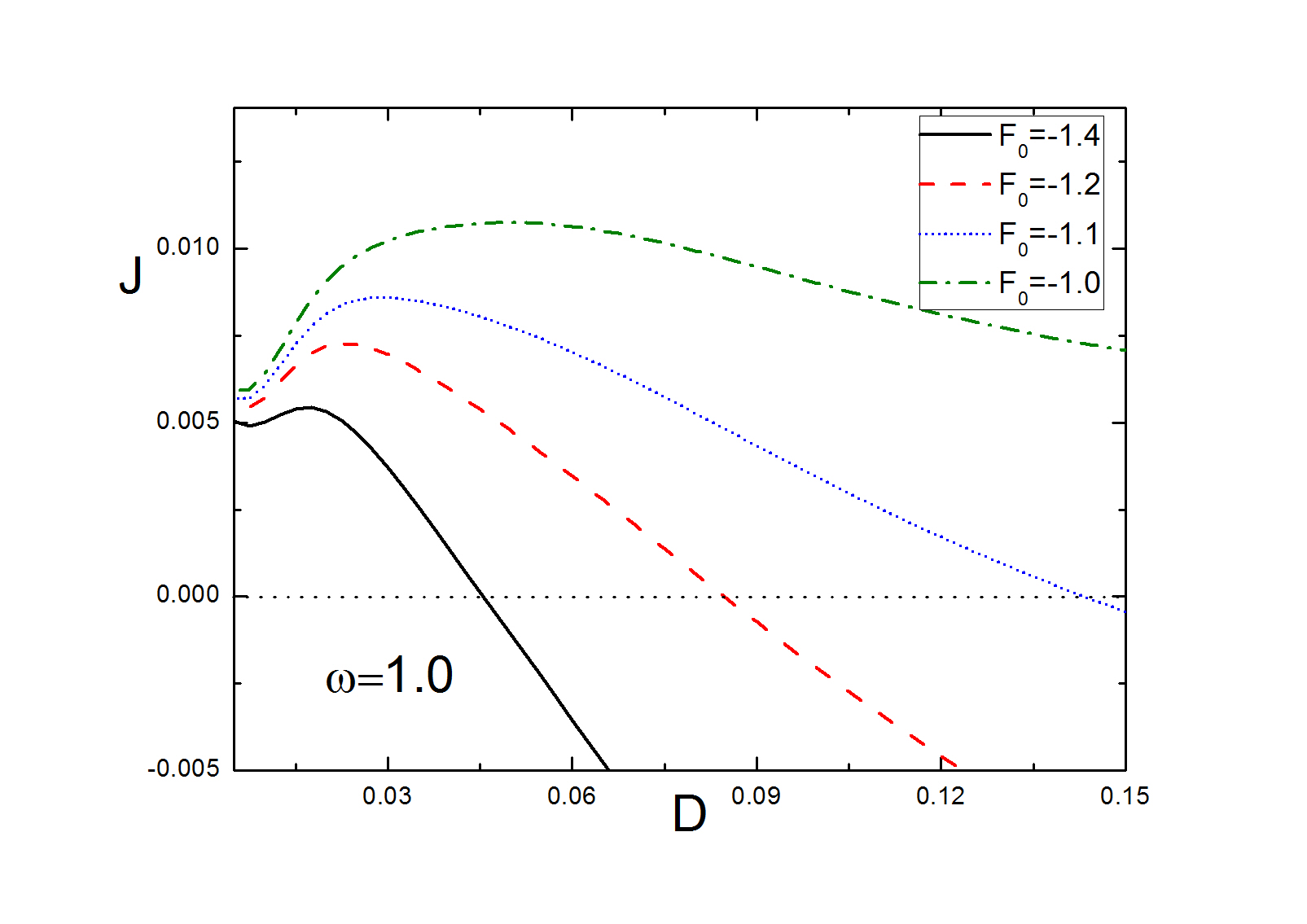}}
\end{minipage}%
\vfill
\caption{J versus $D$ for different values of $F_0$ indicated in the figures. Panel a) corresponds to  $\omega=0.5$ and b) to $\omega=1$. Other parameters as in Fig.\ref{fig:conturs}}
\label{fig:conturs1}
\end{figure}

Numerical results for J are obtained as a function of $D$ and $F_0$. In Fig.\ref{fig:conturs} we present a contour plot of J versus $D$ and the strength of the external force $F_0$ . In Fig.\ref{fig:conturs1} a) and b) we depict some representative curve levels for $\omega=0.5$ and $\omega=1.0$, respectively . We observe the presence of a current reversal (CR) for which J switches from negative to positive values, even for negative values of the external force. Although $F_0 <0 $, the entropic force and the effective diffusion during a time period have an effect that compensate the negative external drift.  Therefore, there is a particle transport in a direction opposed to the force, assisted by entropic contributions.  The current inversion is found for all frequencies in the range considered. 

On the other hand when $F_0$ is tuned, the current as a function of $D$ exhibits monotonic and non-monotonic behaviors. The strictly monotonic behavior depends on the frequency value and is found approximately for $F_0 \leq -1.5$ or $ F_0 \geq -0.5$ (see Fig. \ref{fig:conturs}). For intermediate values of the force, a resonant regime is found where an optimal diffusion coefficient maximizes the transport thus yielding the most favorable situation for the coherent motion of particles in the confined geometry. However, this effect depends on the frequency, being more remarkable as long as $\omega$ increases in which case the maximum moves to higher values of $D$. 

\begin{figure}[h]
\centering
\includegraphics[width=8cm,height=6cm]{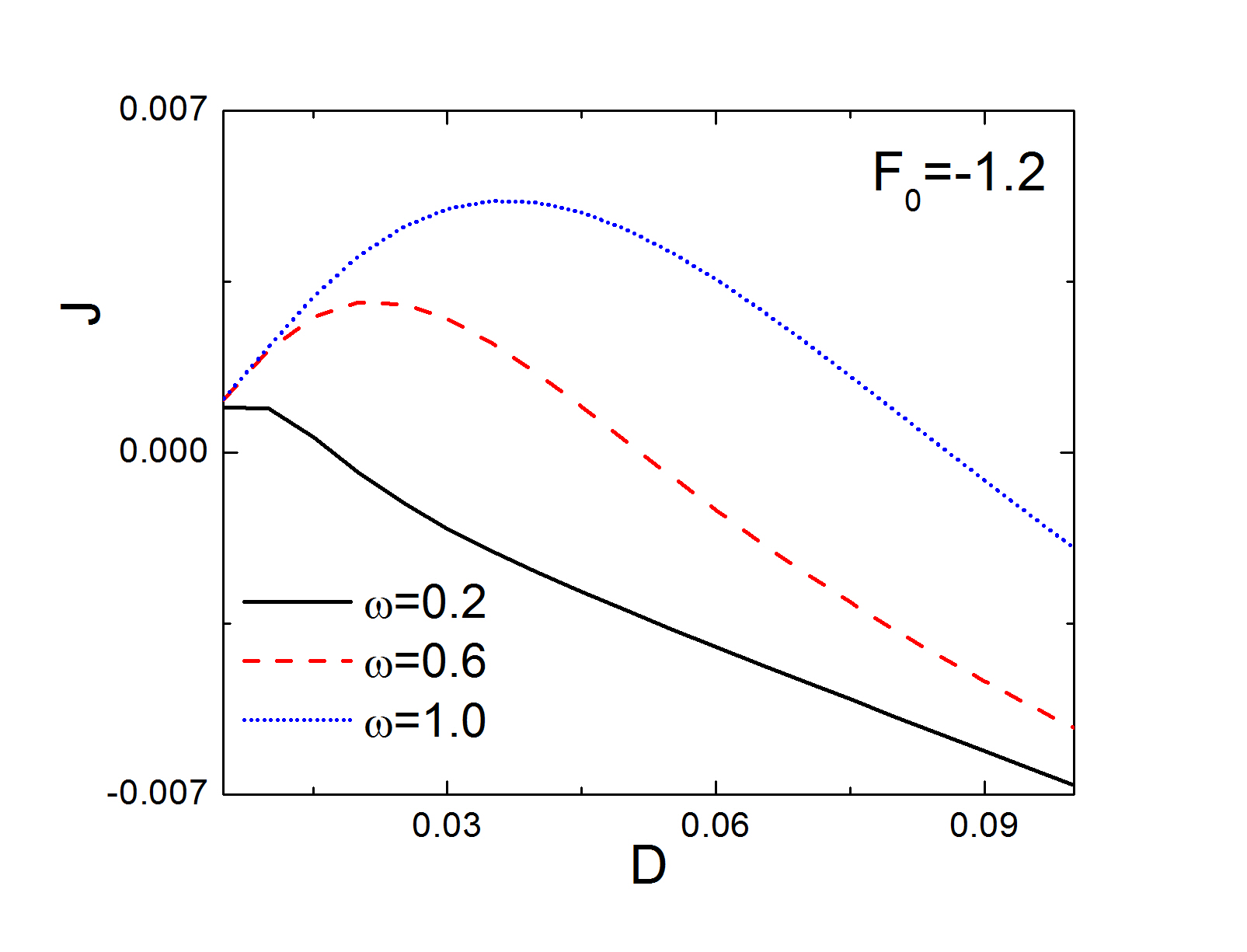}
\caption{J versus $D$ for $\omega =0.1$ (black), $\omega =0.5$ (red) and $\omega =0.8$(blue) and $F_0=-1.3$. The values of the remaining parameters are the ones used in Fig.\ref{fig:conturs}}
\label{JvsD}
\end{figure}

To clarify this point, we compare in Fig.\ref{JvsD} the average current versus $D$ when $F_0$ is fixed but $\omega$ varies. We observe a resonant effect that is more pronounced when the frequency increases. Moreover, the peak also moves to larger values of $D$. CR also occurs for larger $D$ as long $\omega$ increases. 

\begin{figure}[h]
\centering
\includegraphics[width=7cm,height=5cm]{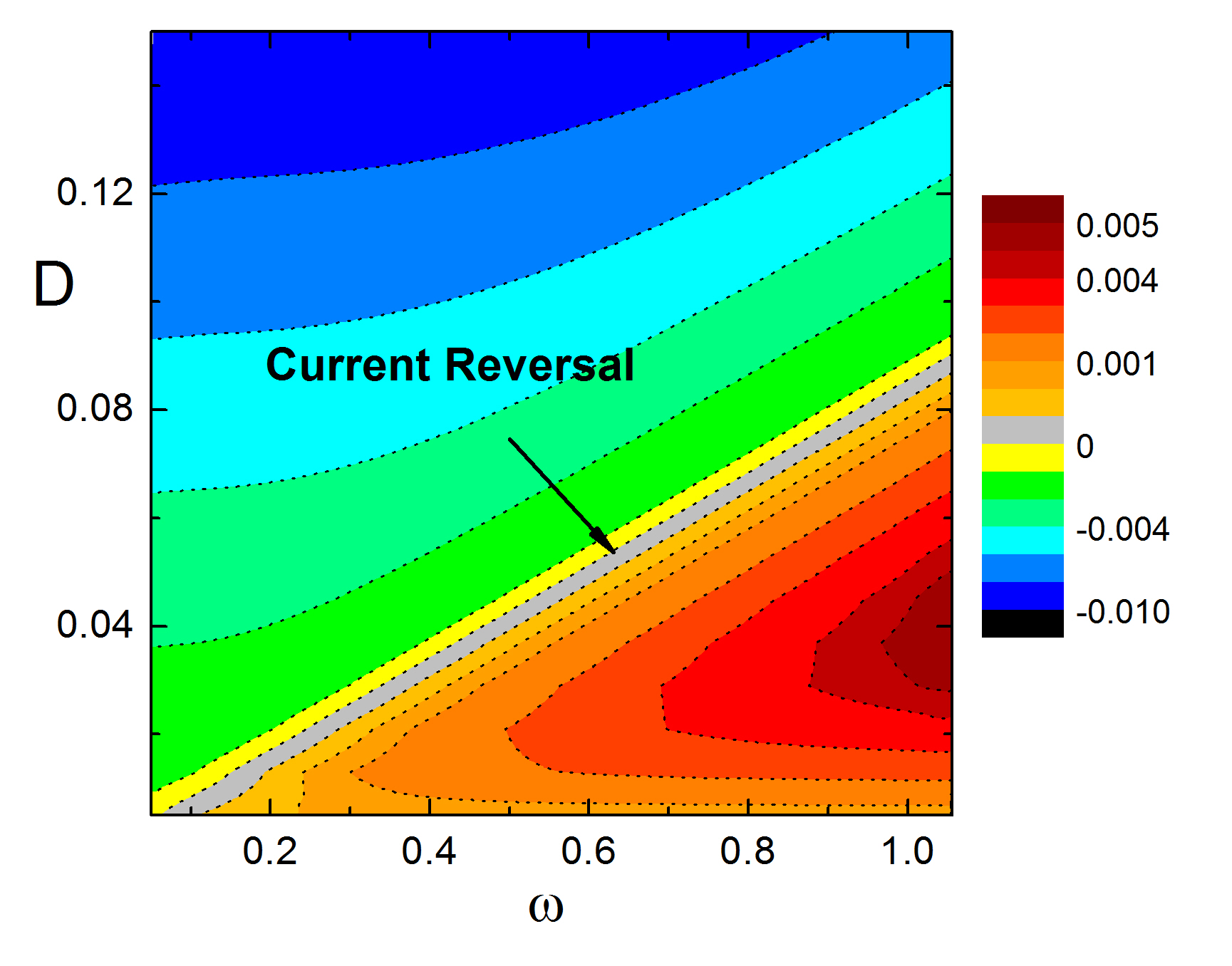}
\caption{Contour plot for J versus $\omega$ and $D$. $F_0=-1.3$ and other parameters as in Fig.\ref{fig:conturs}}
\label{conturW}
\end{figure}

In order to display all these effects together, we show in Fig.\ref{conturW} J in the $D$ - $\omega$ plane. In the figure, we indicate the region where current reversal occurs which separates a region of negative monotonic current from a region of positive current with resonant behavior.

This interesting phenomenon suggests some kind of diffusion-selected effect. Adjusting the frequency of modulation and as $D$ depends on properties of the particles, it is possible to separate particles moving in different directions or in the same direction, when some are faster than others.

\begin{figure}[h]
\begin{minipage}{0.5\textwidth}
\subfloat[]{\includegraphics[width=7cm,height=5cm]{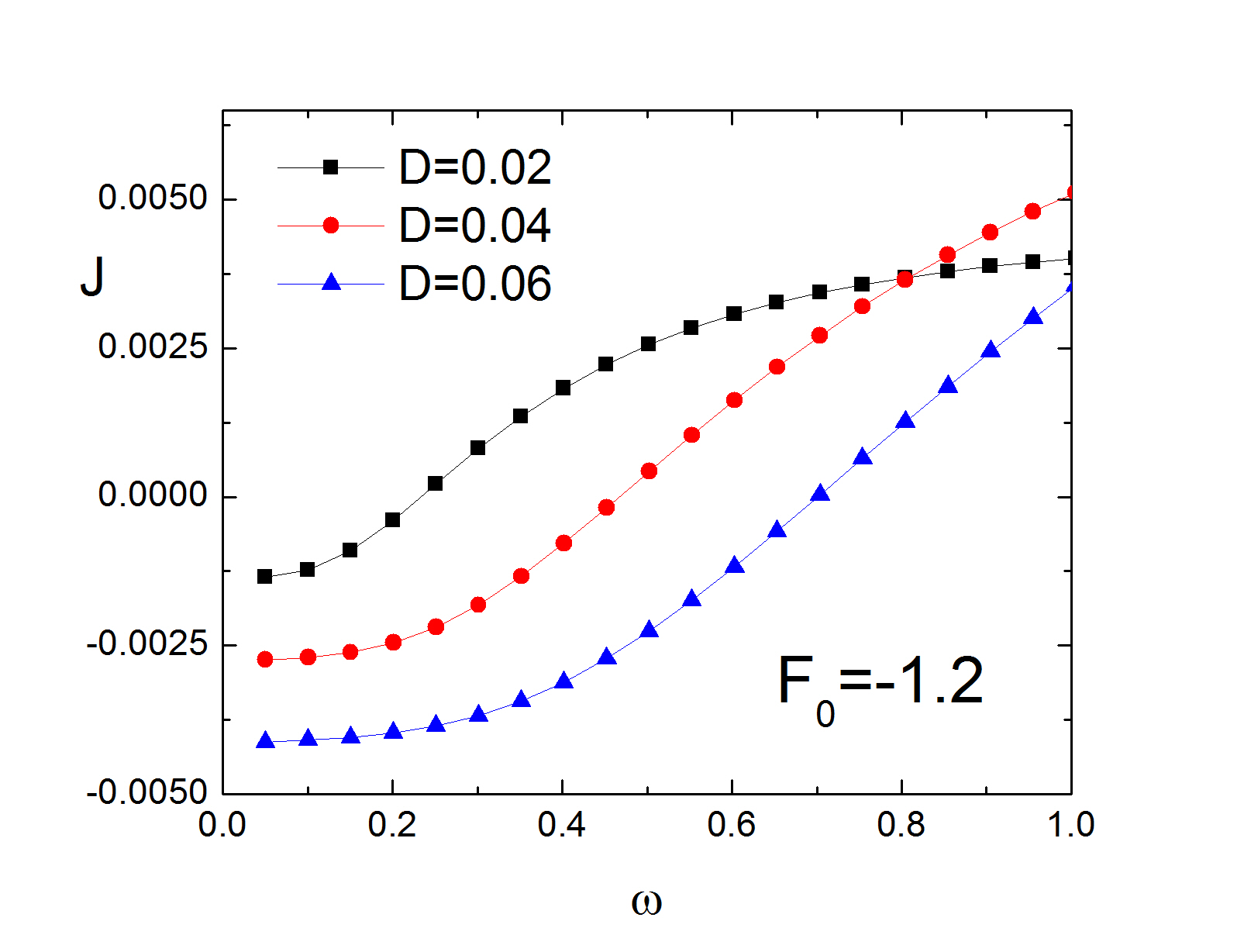}}
\end{minipage}%
\vfill
\begin{minipage}{0.5\textwidth}
\subfloat[]{\includegraphics[width=7cm,height=5cm]{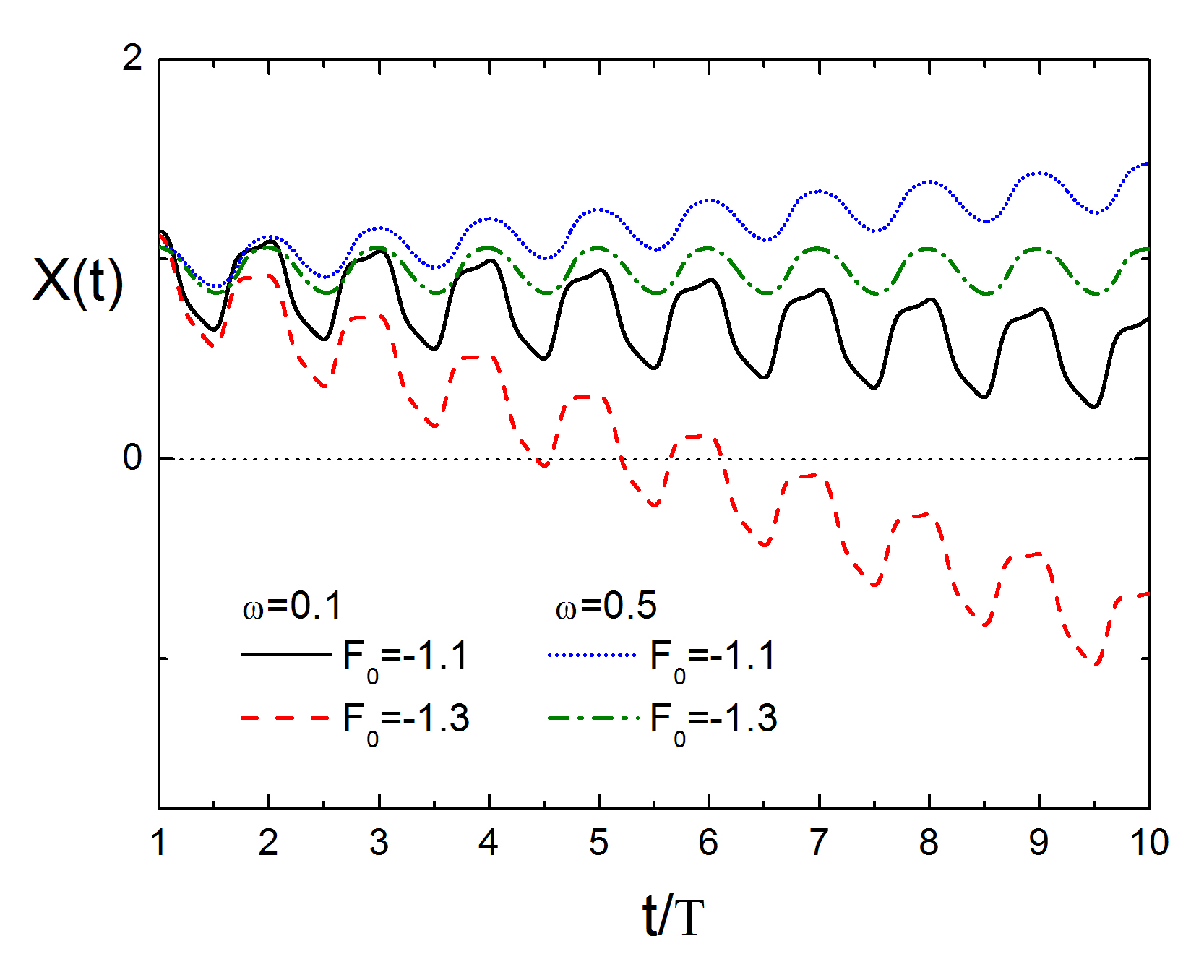}}
\end{minipage}%
\caption{a) J versus $\omega$ for $D=0.02$ (black), $0.04$ (red) and $0.08$(blue) and $F_0=-1.3$. b) $X$ versus $t/T$ for $D=0.03$. }
\label{JvsW}
\end{figure}

It is interesting to understand the role and the relative importance of the two relevant times scales involved in the transport.  One is associated to the frequency of the oscillations and the other to the diffusion process. In Fig. \ref{JvsW}a) the average particle current J is depicted as a function of $\omega$, for different values of $D$. We choose $F_0=-1.2$ such that the possibility of a resonant behaviour depends only on $D$ (see Fig.\ref{JvsD}). We observe that J increases monotonically with $\omega$, irrespective of the value of $D$ . But when diffusion is low enough the current rectification takes place in the same direction than the external force.  Despite J increases with $\omega$, the growth is slow for low diffusion. In all the cases a current inversion is found. 

Many current experimental studies on transport of single or ensembles of colloidal particles are based on tracking techniques by video microscopy. Thus it becomes relevant to study not only velocities but also  average positions and their dependence on time and on the relevant parameters. Moreover, a set up sensitive to specific properties of the colloids could be used as a selective rectifier. In previous works \cite{reguera, Motz} a particle splitter based on a static geometry under the action of time dependent external forces was studied, however in our model the mechanism relying on the modulation of the walls plays a key role on the transport properties observed.

In Fig.\ref{JvsW}b), the mean position of the ensemble of particles versus time is depicted according to Eq.\ref{xt}. Initially, particles are uniformly distributed along one unit and the strength of the external force is taken smaller or close to the force that causes current reversal. It is expected that when $F_0$ increases the current also increases. However, we observe that for a given value of $F_0$ it is possible to increase the drift (i.e. J) just by increasing the frequency. As we showed above, even for a fixed $F_0$ it is possible to rectify the transport in an opposite direction upon variation of $\omega$.  For instance, for $F_0=-1.1$ and low frequency particles move to the left. However increasing $\omega$ the mean position is confined in one unit of the channel during many periods with a small drift to the right. Something similar is observed for $F_0=-1.3$. Increasing frequencies the systems goes from a small and negative current to a positive current, spending many periods in one unit cell.
The center of mass of the ensemble undergoes an oscillatory motion inside one unit when CR occurs.

\begin{figure}[h]
\begin{minipage}{0.5\textwidth}
\subfloat[$F_0=-1.2$ and $\omega =0.1$]{\includegraphics[width=7cm,height=5cm]{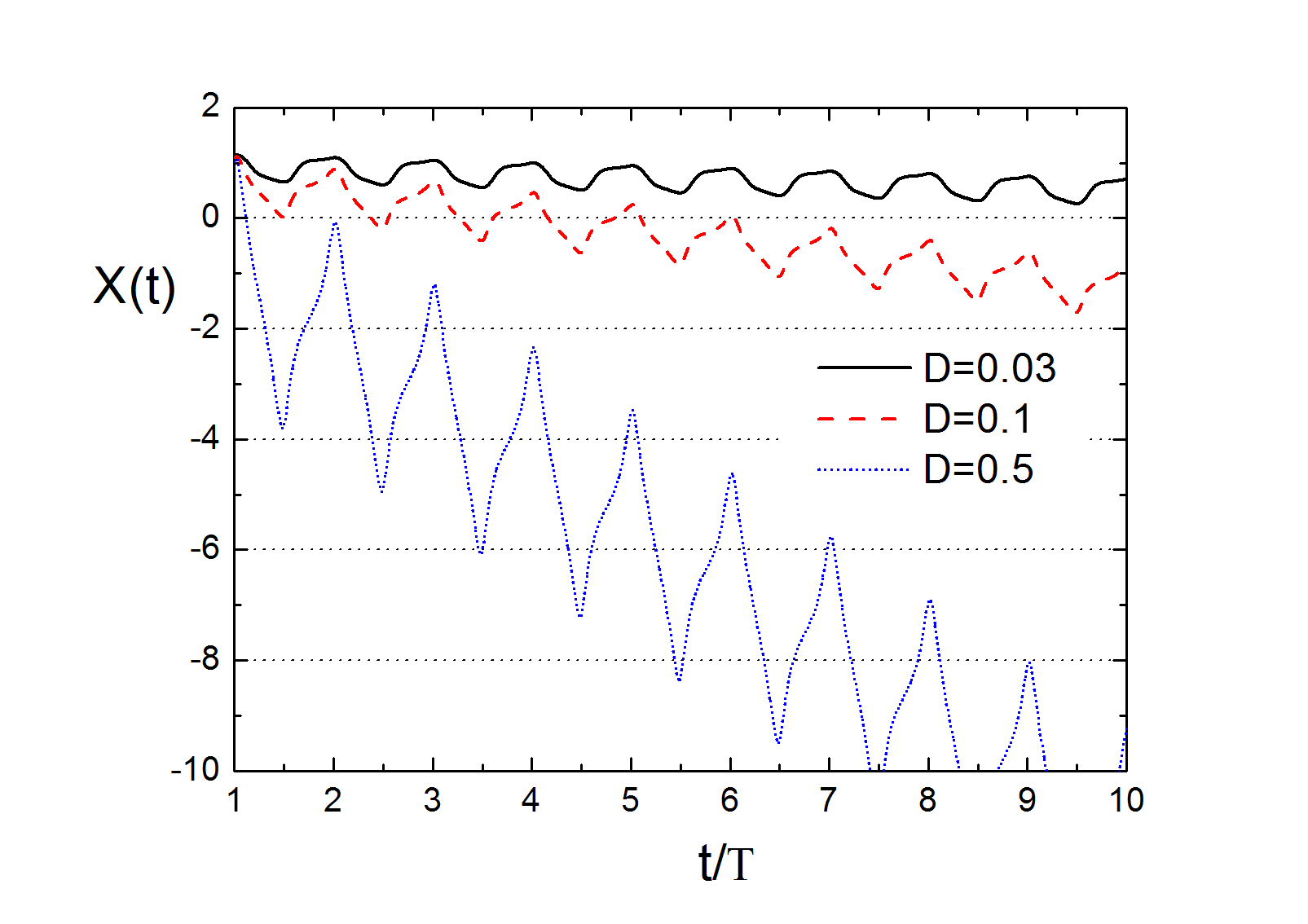}}
\end{minipage}%
\vfill
\begin{minipage}{0.5\textwidth}
\subfloat[$D=0.03$ and $\omega =0.1$]{\includegraphics[width=7cm,height=5cm]{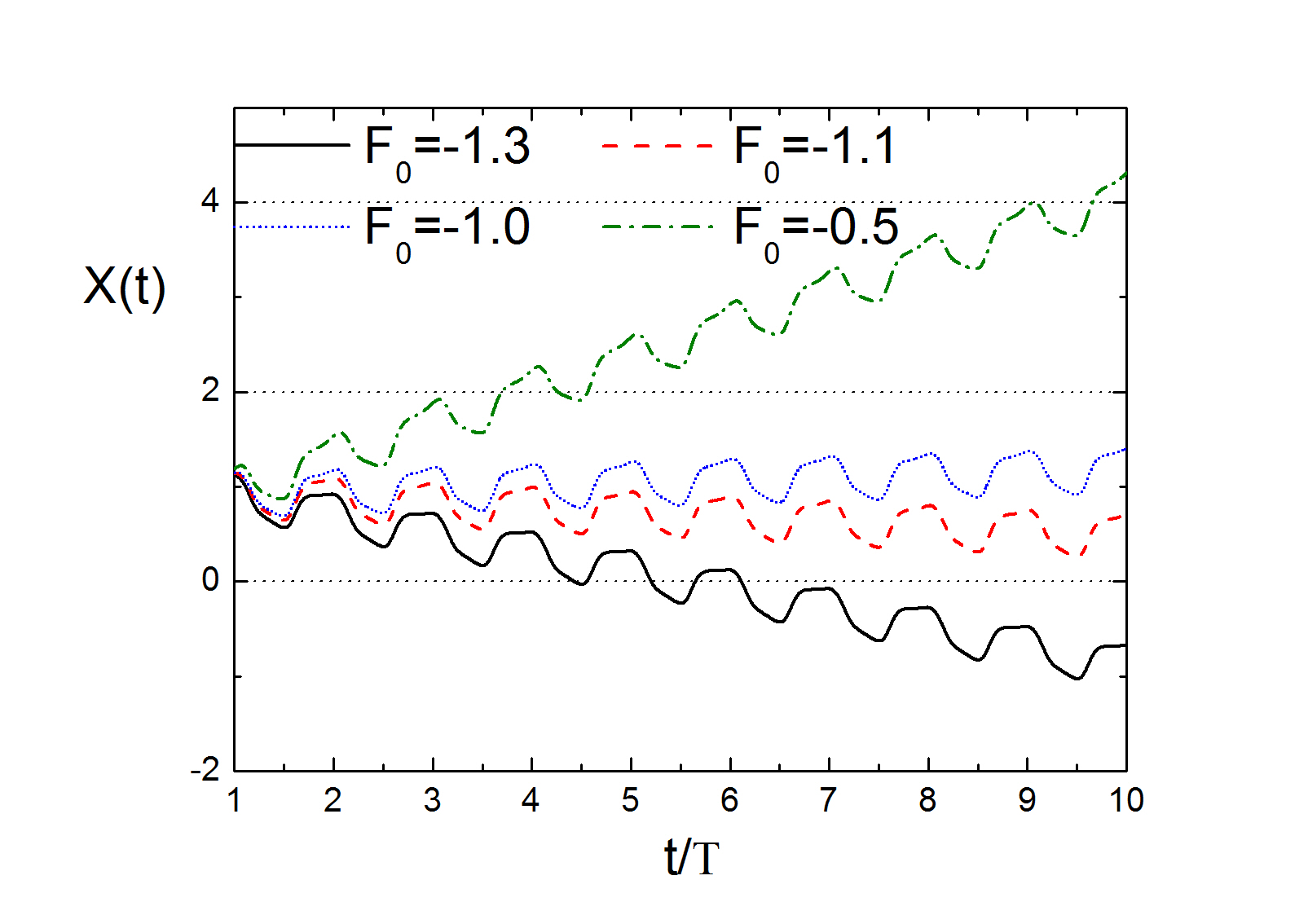}}
\end{minipage}%
\vfill
\caption{$X(t)$ versus $t/T$ for different values of $D$ (a) and $F_0$ (b). Horizontal dot lines separate units of length. The remaining parameters corresponds to the ones in Figure\ref{fig:conturs}}
\label{XvsT}
\end{figure}

In Fig.\ref{XvsT}, we plot the averaged position of the ensemble of particles versus time given in Eq.(\ref{xt}), for $\omega=0.1$ and when $F_0$ or $D$ are kept fixed ( figures (a) and b) respectively).

We observe interesting situations. For instance Fig. \ref{XvsT}a) corresponds to $F_0=-1.2$ (close to CR) and low frequencies. For low diffusion, $X(t)$ is confined in one unit cell. If $D$ increases, the oscillations are still confined during many periods but with a slight drift. In these situation, the motion is mostly bounded during longer times. This property can be useful for a tracking procedure.  For higher diffusion, the motion is still oscillatory but with an important drift. Effective diffusion enhances the drift and the oscillations amplitude. On the other hand, in panel b) it is presented a situation with low $D
$ and $F_0$ tuned around the CR value. As expected, depending on the force strength the evolution of $X(t)$ can be bounded or unbounded with almost constant amplitude of oscillations ($D$ fixed). Again, it is interesting to observe the CR phenomenon where the motion is essentially due to the oscillation of the center of mass in one unit cell, from one side of the bottleneck to the other side.  

\begin{figure}[h]
\begin{minipage}{0.5\textwidth}
\subfloat[$F_0=-1.2$ and $\omega =0.1$]{\includegraphics[width=7cm,height=5cm]{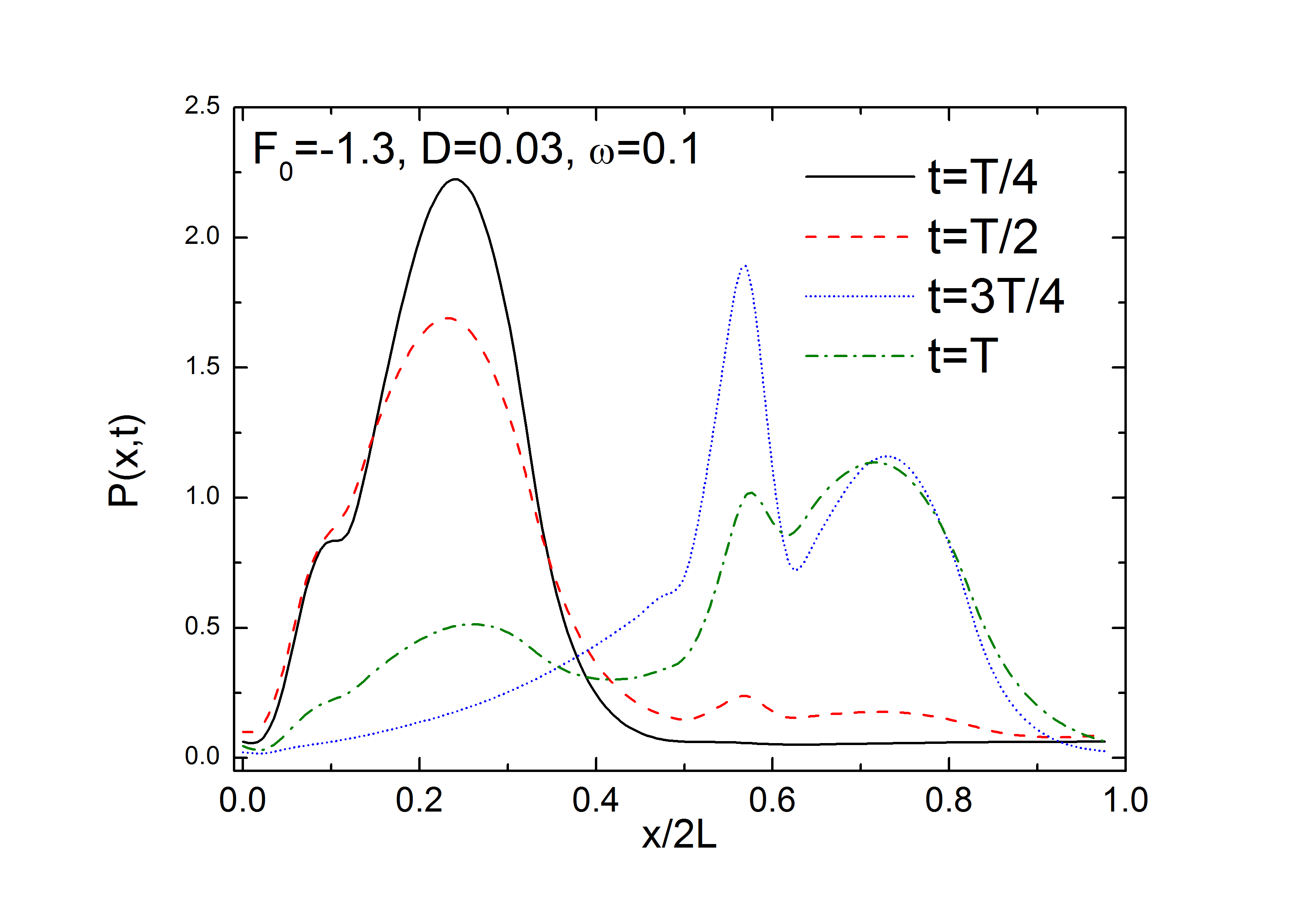}}
\end{minipage}%
\vfill
\begin{minipage}{0.5\textwidth}
\subfloat[$D=0.03$ and $\omega =0.1$]{\includegraphics[width=7cm,height=5cm]{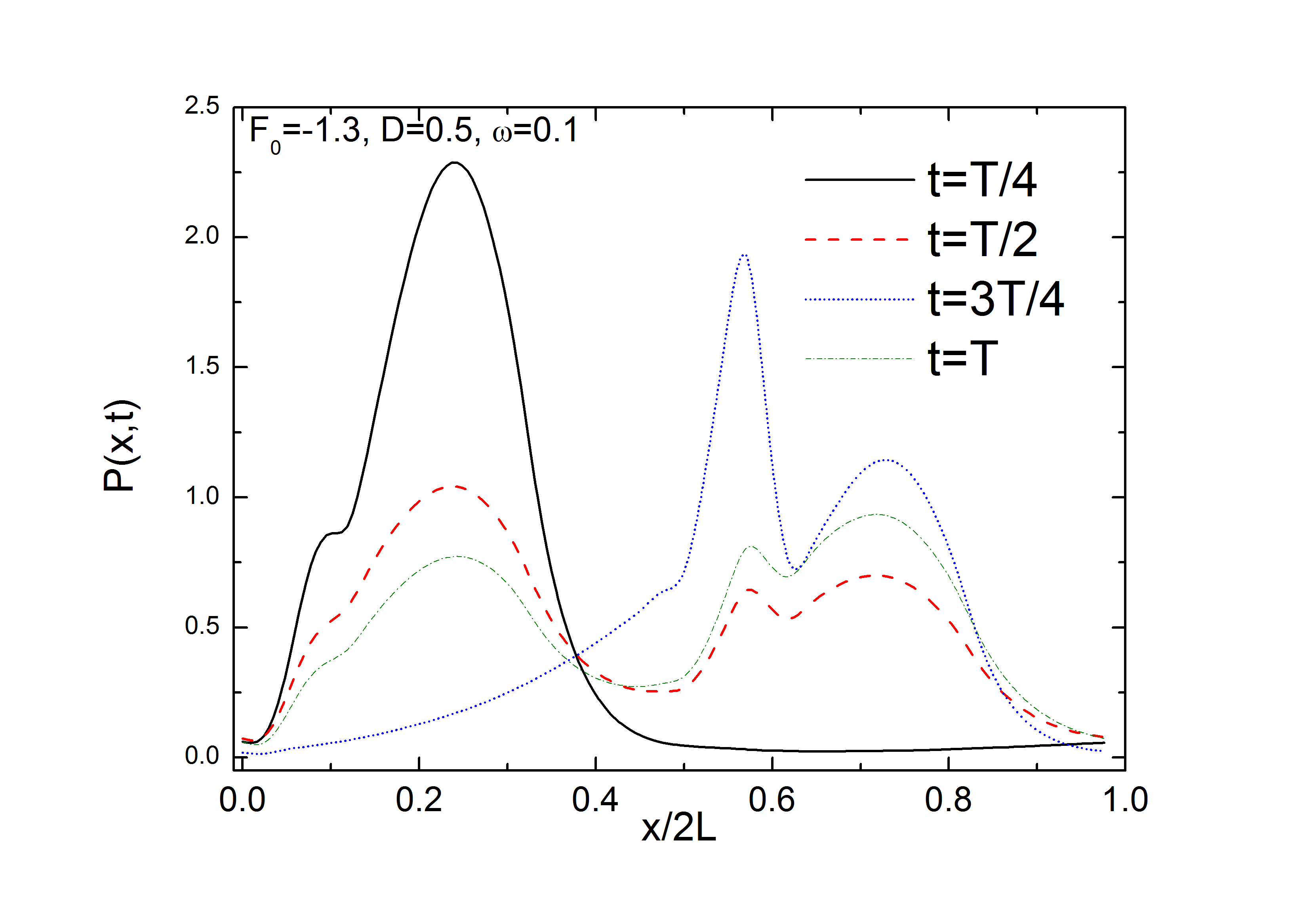}}
\end{minipage}%
\vfill
\caption{Probability distribution $P(x,t)$ for $T/4$ (black full line), $T/2$ (red dashed line), $3 T/4$ (blue dot line) and $T$ (green dashed-dot line). $D=0.03$ (a) and $D=0.5$ (b).  $F_0=-1.3$ and $\omega = 0.1$. Lower panel to $D=0.5$.}
\label{prob}
\end{figure}

In Fig.\ref{prob}, we show the evolution of the probability density $P(x,t)$ at four different times within a period $T$. The figures corresponds to $D=0.03$ and $\omega=0.1$ a) and $0.5$ b) with $F_0=-1.3$.
As expected, the probability distribution oscillates in such a way that the maximun is located at each side of the unit cell alternatively. This explains the fact that the time evolution of the average position of the center of mass, given by the first term of Eq. \ref{xt}, oscillates from one side to the other of the bottleneck. If $D$ is small, all the particles are located more likely in one or other side of the unit cell every half time period. However, increasing diffusion the particles become distributed in both sides longer times. Due to periodic boundary conditions of $P(x,t)$, this situation enhances the net drift given by the second term of Eq.\ref{xt}.

\vspace{1cm}

\begin{figure}[h]
\centering
\includegraphics[width=8cm,height=6cm]{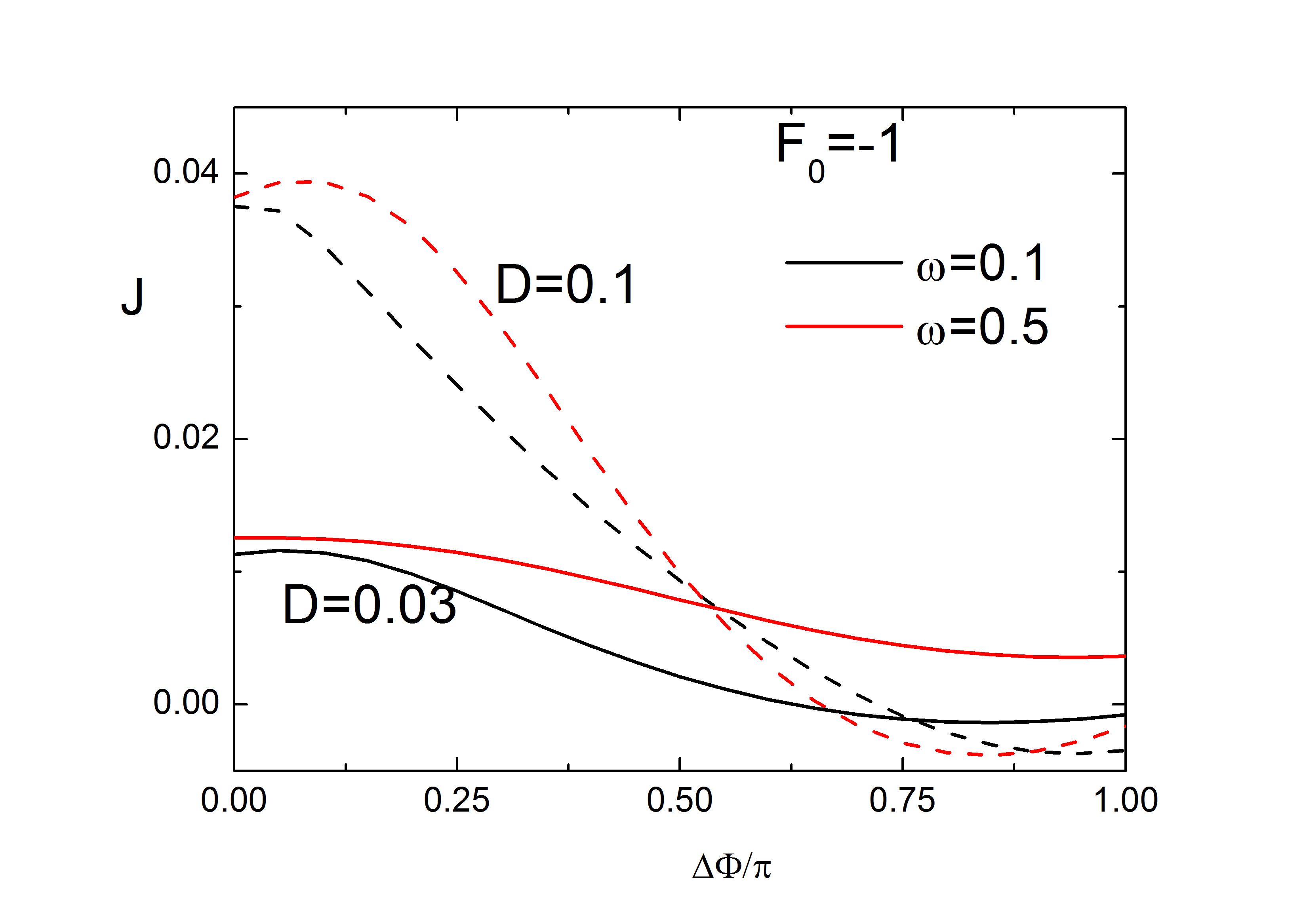}\\
\caption{J versus $\Phi / \pi$ for $D=0.03, 0.1$ and $\omega = 0.1, 0.5$. $F_= -1.0$.}
\label{JvsFi}
\end{figure}

In the previous analysis, we have considered that the fluid in the channel is incompressible. It is also interesting to analyze if the transport through compressible fluids is affected by the presence of the pulsating walls and how it depends on the relevant parameters.  We then analyze the transport for phase differences in the interval $[0,\pi]$.  In Fig.\ref{JvsFi}, we show J as a function of $\Phi$. We observe a current reduction when $\Phi \rightarrow \pi $. Therefore compressibility seems to favor transport and rectification.  However, the dependence on $D$ is non trivial. When the phase lag between adjacent subunits is small, a high diffusion favors the increase of J. Transport is mainly due to the diffusive process and $\tau_{dif}$ is the relevant time scale. Additionally, higher $\Phi$ favors a decrease of the current and J seems to be less dependent on $D$, whatever the value of $\omega$ is. In this limit the dynamical time scale associated to $\omega$ appears to have a comparable relevance as $\tau_{dif}$.

\vspace{1cm}

It is well known that the entropic transport is highly dependent on the asymmetry of the geometry. The shape and parameters used in our work guarantee an asymmetric geometry meanwhile the walls oscillate.  In particular we are interested in the width of the bottleneck $d$ or entropic barrier, because it plays a crucial role related to the flow of particles between adjacent subunits.

\begin{figure}[h]
\centering
\includegraphics[width=8cm,height=6cm]{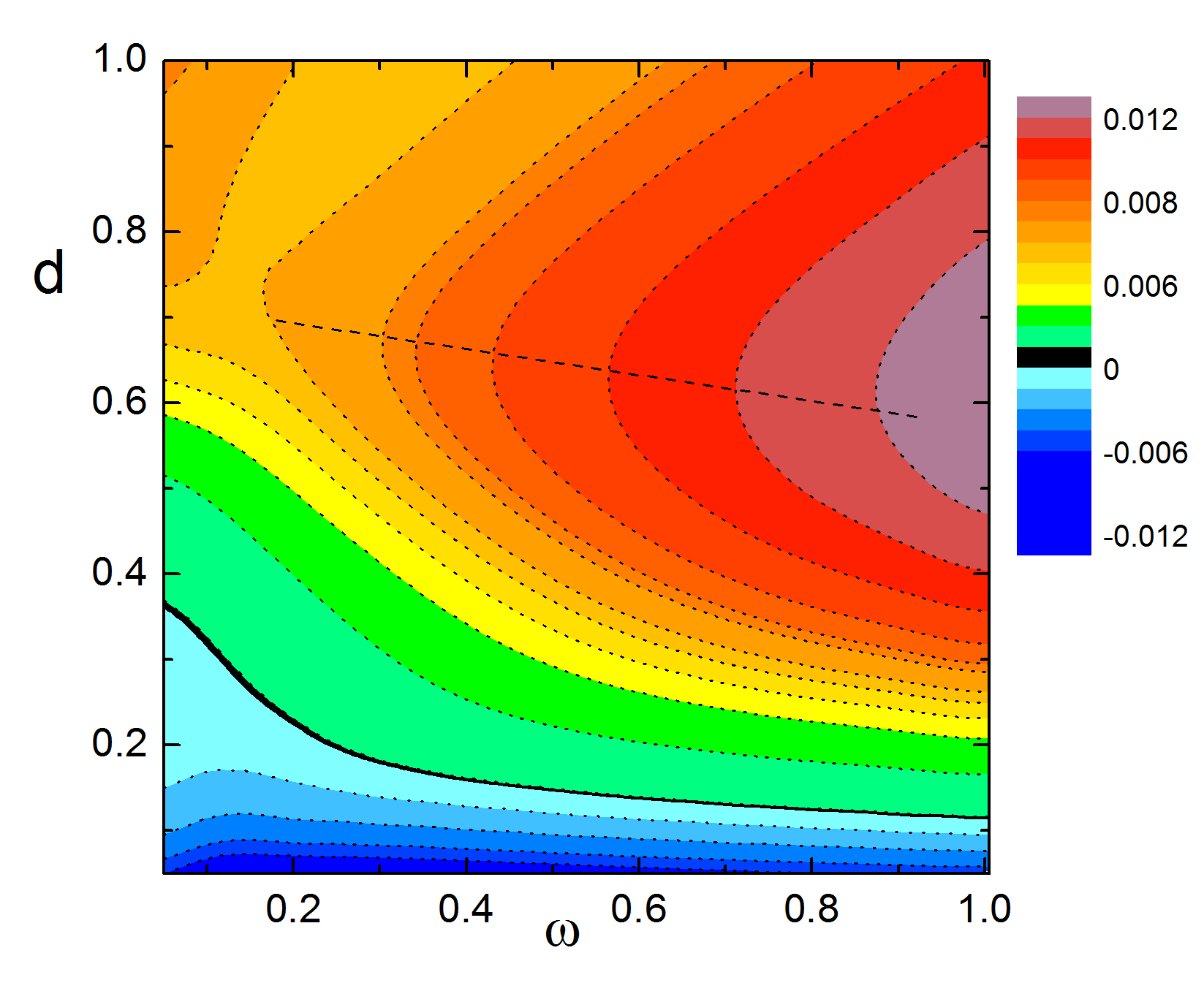}\\
\caption{Average current J in the $d$-$\omega$ plane.  $F_0=-1.1$ and $D=0.03$. The black solid contour line corresponds to current inversion. The dashed line indicates the direction of the maximun.}
\label{Conturd}
\end{figure}

\begin{figure}[h]
\centering
\includegraphics[width=7cm,height=5cm]{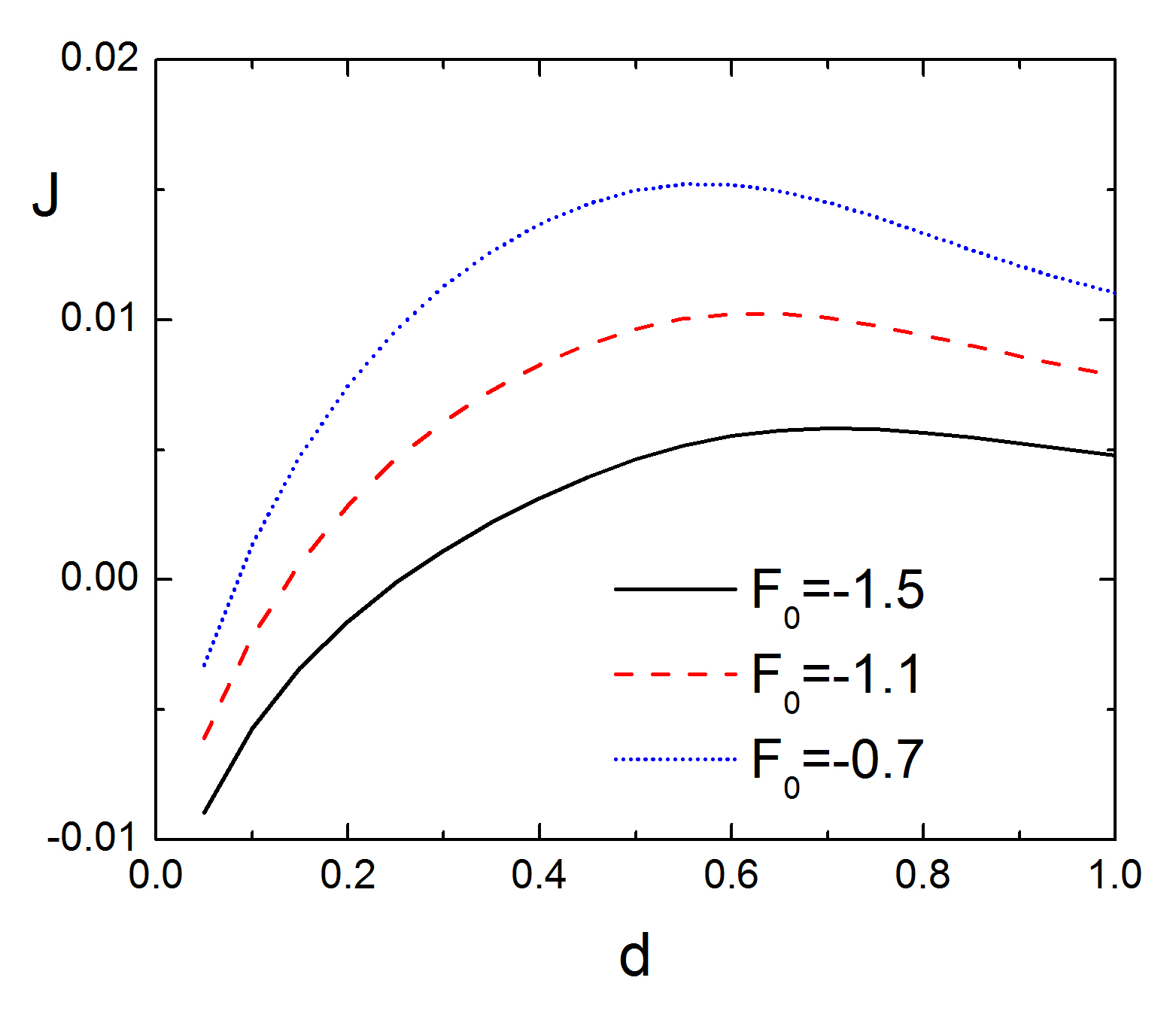}\\
\caption{ J versus $d$ for $F_0=-1.5$ (blue full line), $-1.1$ (red dashed line) and $-0.7$ (blue dotted line).  $D=0.03$ and $\omega = 0.5$.}
\label{Jsd}
\end{figure}

In Fig.\ref{Conturd}, we depict a contour plot of J in the $d-\omega$ plane. For $\omega < 0.2$,  J increases monotonically with $d$ and for frequency $\omega \approx 0.2$ is almost constant for higher $d$. From this $\omega$ a resonant response is found for which the maximum J depends on the frequency.  In Fig. \ref{Jsd} we show the dependence of resonant curves with $F_0$.  Contrary to what one could expect, a wide bottleneck does not enhance transport.  For an oscillation frequency there is an optimal width that optimizes the rectification. The three curves have similar shapes but are shifted suggesting that the optimal $d$ is almost independent of $F_0$. This is a clear evidence that $d$ plays a key role concerning the entropic contribution to transport. Moreover, we observe that the optimal value of $d$ slightly decreases with $\omega$, so for faster oscillations a smaller bottleneck is required in order to optimize transport.

\section{Conclusions}

We have investigated the dynamics of Brownian particles in a confined channel whose shape is periodically modulated in time. 
We have identified a particle current rectification regime. Rectification has also been found in the direction opposite to the external force, where entropic contribution plays a key role. The average particle current changes with $D$ monotonically, or non-monotonically showing a resonant behavior.  In the last case, the maximum is strongly dependent on $F_0$ and $\omega$ and shifted to higher values of $D$ when walls oscillate faster.
Moreover, we found a current reversal that takes place when negative external forces are acting. The phenomenon observed requires the time-dependent entropic barrier that affects the force and the diffusion.

For incompressible fluids transport is less efficient but more favorable for the appearance of CR, provided that $\Phi$ is sufficiently small and is enhanced for larger $D$ . On the other hand the more realistic limit consisting in an incompressible fluid, is less beneficial to increase velocity but it favors the appearence of CR if $D$ or $F_0$ are properly chosen.

Recent works considered the case of directional transport due to time-dependent forces and static boundaries \cite{rubi5} or time-dependent boundary and constant diffusion \cite{entropic,entropicw}. Our model not only includes a time dependent boundary but also its effect on an effective time-dependent diffusion.  We found that it is possible not only to rectify the transport but also to optimize and invert its direction, phenomena that appears as a transition between an energy-dominated regime and a entropic-dominated one. 

Our results show the constructive role played by time-modulated boundaries in the transport properties of particles.  This fact can be interesting in many practical situations as in the rectification of microswimmers motion, Janus particles, substances delivery or vascular systems, to mention just some few. The pulsating action required  could be exerted by externally-controlled actions or by the system itself, processes which may be of great significance in physico-chemical or biological systems taking place in confined geometries.

\section{Acknowledgments}

This work is supported by CONICET-Argentina under Grant No. 14420140100013CO and MICINN of the Spanish Government under Grant No. FIS2015-67837-P.

\end{document}